\documentclass[prl,aps,superscriptaddress,twocolumn]{revtex4}
\usepackage{graphicx}
\usepackage{amsmath}
\usepackage{amsfonts}
\usepackage{amssymb}
\usepackage{epsfig}
\usepackage{color}
\usepackage{dsfont}

\newcommand{\numberset}{\mathbb}

\newcommand{\Z}{\numberset{Z}}

\newcommand{\abs}[1]{\left\vert#1\right\vert}

\newcommand{\bra}[1]{\left\langle#1\right\vert}
\newcommand{\ket}[1]{\left\vert#1\right\rangle}
\newcommand{\Tr}[1]{\text{Tr}\left\{#1\right\}}

\begin{document}

\title{Topological superconducting phases and Josephson effect in curved time-reversal-invariant superconductors}

\author{G.~Francica}
\affiliation{CNR-SPIN, c/o Universit\`a di Salerno, I-84084 Fisciano (Salerno), Italy}

\author{M.~Cuoco}
\affiliation{CNR-SPIN, c/o Universit\`a di Salerno, I-84084 Fisciano (Salerno), Italy}
\affiliation{Dipartimento di Fisica ``E. R. Caianiello'', Universit\`a di Salerno, I-84084 Fisciano (Salerno), Italy}

\author{P.~Gentile}
\affiliation{CNR-SPIN, c/o Universit\`a di Salerno, I-84084 Fisciano (Salerno), Italy}
\affiliation{Dipartimento di Fisica ``E. R. Caianiello'', Universit\`a di Salerno, I-84084 Fisciano (Salerno), Italy}


\begin{abstract}
We consider a Rashba spin-orbit coupled nanowire with anisotropic spin-singlet superconducting pairing and time-reversal-invariant symmetry. We explore the evolution of the topological superconducting phases of this system due to geometric deformations for the representative case of a wire bent in a semielliptical shape. We find that when the system is in its topological superconducting phase, strong inhomogeneities in the profile curvature can produce a pair of localized eigenmodes, which can be attributed to a nonuniform topological phase.  
The curved geometric profile also allows to tune the spin correlations of the superconducting state via the induced inhomogeneity of the spin-orbit coupling (SOC). The geometric control of the superconducting pair correlations allows to manipulate the critical current in Josephson junctions made up of two time reversal invariant topological superconductors separated by a spin-orbit coupled normal metal. In particular, we find that the curvature inhomogeneity can be exploited for amplifying  the current intensity, but also to generate a $0-\pi$ transition, and a second harmonic contribution, which generates, for some specific geometric configurations, a $\varphi$-junction behavior.
\end{abstract}

\maketitle

\section{I. Introduction} 
Topological superconductors (TSCs) have recently attracted a lot of experimental and theoretical attention due to the possibility to generate Majorana modes at the ends of one-dimensional TSC ~\cite{read00,kitaev01,sato17,qi2011}. Majorana modes are topologically protected, and are particularly attractive for realizing quantum registers which are immune to decoherence effects~\cite{kitaev01}, with promising applications in fault-tolerant quantum computation ~\cite{nayak2008}.

Recent proposals have predicted the possibility to realize topological superconductivity by hybridizing ordinary superconductors (SCs) with helical materials, with the help of magnetic perturbations ~\cite{fu&kane2008, sato2009, sau2010,alicea2010, lutchyn2010,oreg2010}. Due to proximity, electrons in a single helical band at the Fermi energy realize a spinless chiral $p$-wave SC in its weak pairing regime, corresponding to a topological SC with broken time reversal symmetry (class D).  The presence of Majorana modes in this class of TSCs is associated with unique and interesting physical signatures, including zero bias conductance peaks, anomalous Fraunhofer patterns, and fractional Josephson effects  ~\cite{flensberg2010,mourik12,deng2012, vel2012, will2012, rok2012,leijnse12}. 

More recently, a completly distinct family (class DIII) of time-reversal invariant TSCs has been proposed, based on a mathematical classification of Bogoliubov-de Gennes (BdG) Hamiltonians ~\cite{classif2008,qi2009, teo2010,schn2010}. In these TSCs the zero modes appear in pairs due to Kramers theorem, differently from the chiral TSCs. As a result, two Majorana bound states (MBSs) will be localized at each end of the DIII-TSC nanowire, leading to the formation of one Kramers doublet ~\cite{zhang2013, nakosai2013}.  
It is interesting to note that, even though two local MBSs together form a usual fermion, the exchange of two Kramers pairs of MBSs constitutes a non-Abelian operation \cite{naMBS}. 

Up to now, theoretical groups have suggested many proposals to realize the DIII-TSCs in hybrid structures based either on superconductors with $d_{x^2-y^2}-$wave ~\cite{wong12} or $s\pm$-wave ~\cite{zhang2013} pairing, noncentrosymmetric superconductors ~\cite{nakosai2013}, bilayers of two-dimensional superconductors with spin-orbit coupling ~\cite{nakosai12}, and also in the case of superconductors with conventional $s$-wave pairing ~\cite{deng12, kese13, gaida14, haim14, zhang14,dimitrescu14,loss14}. The case of $s$-wave pairing, explored in particular in a two band model\cite{deng12,kese13} or in hybrid structures including two $s-$wave superconductors \cite{kese13}, has shown that a DIII TSC emerges under the assumption of a $\pi$ phase difference between two pairing potentials in the two subsystems, thus mimicking the $s\pm$-pairing considered in Ref.~\cite{zhang2013}.

Since the Kramers doublet is protected by the time-reversal symmetry, it is expected to inevitably drive some new transport phenomena, quite different from the single Majorana zero modes.
Physicists have also explored the quantum transport phenomena contributed by Majorana doublet in Josephson junctions involving an indirect coupling between DIII class TSCs and $s$-wave superconductors, and some important results have been reported ~\cite{chung2013,Liu2014,Gong2016}.
For instance, in the Josephson junction formed by the DIII class TSCs with Majorana doublets, the period of the Josephson current will be varied by the change of fermion parity in the system ~\cite{Liu2014}. 
More importantly, the presence of Majorana doublet in the DIII class superconductor renders the current finite at zero phase difference, with its sign determined by the fermion parity of such a junction. 
This testifies the interesting and nontrivial role of Majorana doublets in contributing to the Josephson effect. 

An ubiquitous ingredient in all proposals for the realization of topological states, both in topological insulators as well as in TSCs, is Rashba spin-orbit coupling (RSOC) ~\cite{rashba}. It is associated with inversion symmetry breaking and is responsible for the electron spin-momentum locking ~\cite{locking}. Such interaction, originally introduced within the context of two-dimensional semiconductors, over the past 30 years has inspired a vast number of predictions, discoveries and innovative concepts far beyond semiconductors, based on the possibility it allows to manipulate the spin orientation by moving electrons in space, and viceversa, to control electron trajectories via the electron spin \cite{manchon}. 
Recently, it has been recognized that a new ingredient, the non trivial curvature, when combined with RSOC in low-dimensional nanostructures, can act a crucial role in establishing a deep connection between electronic spin textures, spin transport properties, and the nanoscale shape of the system. A non-flat geometry can thus provide novel tools for an all-geometrical and electrical control of the spin orientation \cite{nostroPRB16}, of the quantum structure of the superconducting state \cite{nostroSC}, also including the possibility to induce topological nontrivial phases \cite{nostroPRL15, nostroCM} and promoting new topological effects \cite{niccolo}.
Within the superconducting context, it has been already clarified that in non flat systems, the spatial variation of the Rashba field, through the curvature of the nanostructure, can yield either a local enhancement or a suppression of a local spin-singlet superconducting order parameter \cite{nostroSC}. Furthermore, it has been demonstrated that the geometric curvature can effectively act as a spin torque on the electron spin pairs, allowing to geometrically drive the spatial textures of the spin-triplet pairing, in such a way that the associated $\mathbf{d}$ vector follows the evolution of the electron spin orientation in the normal state \cite{nostroSC}.

In this paper we analyze time-reversal symmetric nanowires which are characterized by RSOC and spin-singlet pairing, with the aim to unveil the effect of the geometric curvature in tailoring a DIII topological phase. We use as prototype geometry an elliptically deformed nanoring, such that, by modifying the ratio between the ellipse minor and the major axes, it is possible to tune the local curvature at will. 
We find that an inhomogeneity in the nanowire curvature can give rise to a Kramers pair of localized quasi-particles, which can be ascribed to a local inhomogeneity of the topological phase. 
Furthermore, the superconducting state displays spin-triplet correlations which strongly depend on the profile curvature. 
These curvature effects are then exploited for manipulating the current in TSC/SOC-semiconductor/TSC Josephson junctions. We analyse how Josephson current can be modulated through a local electrical control, showing that a $\pi$-shift can be achieved by selecting adequately the spin-channels, and a second harmonic can be generated which should give a fractional $\varphi$-junction. Such fractional behavior has been originally predicted in interfaces including unconventional superconductors \cite{larkin,Yip}, and then also found in the context of time reversal symmetry breaking \cite{sigristPC, sigrist98,buzdin08}, in Josephson junctions involving noncentrosymmetric superconductors \cite{NCS2014} and, more recently in junctions between superconductors with different pairing symmetries. In the latter case a $\pi/2$-Josephson junction behavior, with spontaneous time reversal symmetry breaking has been theoretically derived \cite{yang18}.

The paper is organized as follows. In Sec.II we introduce the model system; in Sec III we present the topological properties of the semielliptical nanowire; in Sec. IV we show the curvature-induced spin-triplet correlations; in Sec. V we discuss the Josephson effect in TSC/SOC-semiconductor/TSC junctions in the presence of non trivial curvature, considering and comparing two different geometrical setups. Finally, we give some conclusions in Sec. VI. In the Appendix we discuss the robustness of the topological phase as a function of the model parameters for a finite size system.

\section{II. The model} We consider a superconducting planarly curved wire with a Rashba spin-orbit interaction produced by a transverse electric field pointing along the $z$ direction. The shape of the wire can be specified by introducing two unit vectors $\hat{\mathcal T}(s)$ and $\hat{\mathcal N}(s)$, which are locally tangent and normal to the spatial profile at each space position, labelled by the curvilinear coordinate $s$.
The normal and tangent unit vectors can be expressed in terms of a polar angle $f(s)$ as $\hat  N(s) = (\cos f(s),\sin f(s),0 )$ and $\hat T(s) = (\sin f(s),-\cos f(s),0 )$, and are related each other via the Frenet-Serret type equation $\partial_s \hat N = \kappa \hat T$, with $\kappa = - \partial_s f$ being the local curvature. By adopting a discrete lattice representation \cite{nostroPRB16, nostroSC, nostroPRL15}, and considering both an on-site spin-singlet superconducting interaction of amplitude $\Delta_0$ and a spin-singlet nearest neighbor pair interaction of amplitude $\Delta_1$, the Hamiltonian of the system reads
\begin{eqnarray}
\nonumber H &=& \sum_{i, \sigma\sigma'} \left(c^\dagger_{i\sigma} (-t \delta_{\sigma \sigma'} + \alpha^{\sigma\sigma'}_{i,i+1}) c_{i+1\,\sigma'} +h.c.\right)\\
\nonumber && - \mu \sum_{i,\sigma} c^\dagger_{i\sigma}  c_{i\sigma} + \Delta_0\sum_i (c_{i\,\uparrow}^\dagger c_{i\,\,\downarrow}^\dagger+h.c.)\\
\nonumber &&  + \Delta_1 \sum_{\langle i j\rangle} (c_{i\,\uparrow}^\dagger c_{j\,\,\downarrow}^\dag+h.c.)\\
 &=&  \frac{1}{2} \left[
                                  \begin{array}{c}
                                    \mathbf{c} \\
                                    \mathbf{c}^* \\
                                  \end{array}
                                \right]^\dagger \mathbf{H}_{BdG} \left[
                                  \begin{array}{c}
                                    \mathbf{c} \\
                                    \mathbf{c}^* \\
                                  \end{array}
                                \right]
\end{eqnarray}

\noindent where $\mathbf{c}^\dagger=(c^\dagger_{1\uparrow},c^\dagger_{1\downarrow},\cdots,c^\dagger_{N\downarrow})$, with $c^\dagger_{j\sigma}$ ($c^\dagger_{j\sigma}$) operators creating (annihilating) an electron at the $j$th site with spin projection $\sigma=\uparrow,\downarrow$ along the $z$ axis, $\mu$ is the chemical potential, $t$ is the hopping amplitude between nearest-neighbor sites, and the spin-dependent nearest neighbor hopping amplitudes are
\begin{equation}
\alpha_{j,j+1} = i  \frac{\alpha}{2} (g^x_j \sigma_x + g^y_j \sigma_y)
\end{equation}
where $\alpha$ is the RSOC amplitude, $g^x_j = \cos f_{j}+\cos f_{j+1}$, $g^y_j = \sin f_{j}+\sin f_{j+1}$ , and $f_j$ is the value of the polar angle $f$ at the $j$th site.
The BdG matrix $\mathbf{H}_{BdG}$ can be diagonalized by performing a Bogoliubov transformation $\boldsymbol{\eta} =  \mathbf{u}\mathbf{c} + \mathbf{v}^* \mathbf{c}^*$, so that the Hamiltonian can be expressed in its canonical form $H = \boldsymbol{\eta}^\dagger (\boldsymbol{\Lambda} - \frac{1}{2} )\boldsymbol{\eta}$, where $\boldsymbol{\Lambda}$ is the diagonal semi-definite positive matrix of the single-particle energies~\cite{colpa79}.

\section{III. Topological properties of semielliptical nanowires} 
The system is described by a one-dimensional BdG model characterized by particle-hole and time reversal (TR) symmetry~\cite{chiu16}. In details, the particle-hole and the TR symmetries are represented by two antiunitary operators $\mathbf{C}$ and $\mathbf{T}$, such that $\{\mathbf{H}_{BdG},\mathbf{C}\} =0$, $\mathbf{C}^2=1$ and $[\mathbf{H}_{BdG},\mathbf{T}] =0$, $\mathbf{T}^2=-1$. From these symmetries a chiral symmetry can be built, represented by the unitary operator $\boldsymbol{\Gamma}=\mathbf{C}\mathbf{T}$ such that $\{\mathbf{H}_{BdG},\boldsymbol{\Gamma}\} =0$, $\boldsymbol{\Gamma}^2=-1$. The particle-hole and the time-reversal symmetries can be represented as $\mathbf{C}=\tau_x K$, $\mathbf{T} = i \sigma_y K$, where $K$ is the complex conjugation operator.
Due to these symmetries, the model can be classified as belonging to the DIII Altland-Zirnbauer topological class~\cite{zirnbauer97} and displays homotopic features characterized by a $\Z_2$ topological invariant. In its topological phase the system is expected to have two pairs of zero energy Majorana modes, which are spatially localized at its edges~\cite{chiu16}.

For a straight geometry, the BdG matrix $\mathbf{H}^{straight}_{BdG}$ is translationally invariant, and the topology can be characterized in the momentum space~\cite{schnyder08,kitaev09}. Thus the system is in its topological phase when $\abs{\Delta_0} < 2 \abs{\Delta_1} \land \alpha>\alpha^*$ with $\alpha^* = \abs{t\Delta_0-\mu \Delta_1}/\sqrt{4 \Delta_1^2-\Delta_0^2}$~\cite{zhang13}.

Conversely, for curved structures,  the RSOC becomes non-homogeneous and the physical system becomes aperiodic. 
In the case of a homogenous curvature 
the translational symmetry of the BdG matrix $\mathbf{H}^{homo}_{BdG}$ can be restored by performing the gauge transformation
 \begin{equation}
\tilde c_{j\,\uparrow}= e^{\mathrm{i} j a/2R}  c_{j\,\uparrow}\,,\quad \tilde c_{j\,\downarrow} =e^{-\mathrm{i} j a/2R} c_{j\,\downarrow}
\end{equation}
\noindent where $\mathrm{i}$ is the imaginary unit, $j$ labels the lattice site position,  $a$ is the lattice constant and $1/R$ is the curvature. Within this representation, in the thermodynamic limit $N\to \infty$ we get $\tilde{\mathbf{H}}^{homo}_{BdG}\to \mathbf{H}^{straight}_{BdG}$, so that we can conclude that in the case of a perfectly symmetric ring the curvature does not affect the topological properties of the system.

\begin{figure}[!ht]
\includegraphics[width=0.7\columnwidth, angle=0]{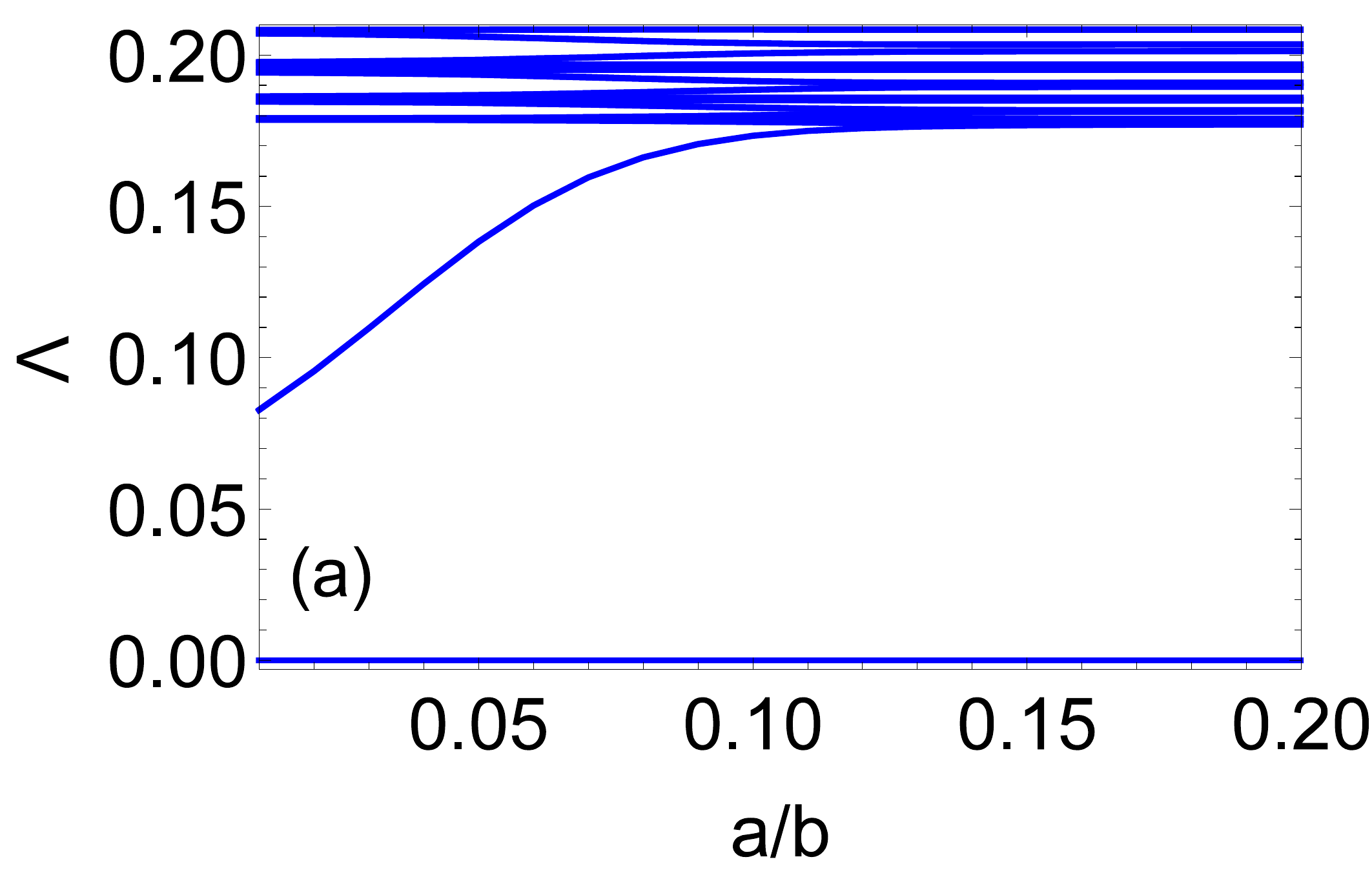}\\ \includegraphics[width=0.7\columnwidth, angle=0]{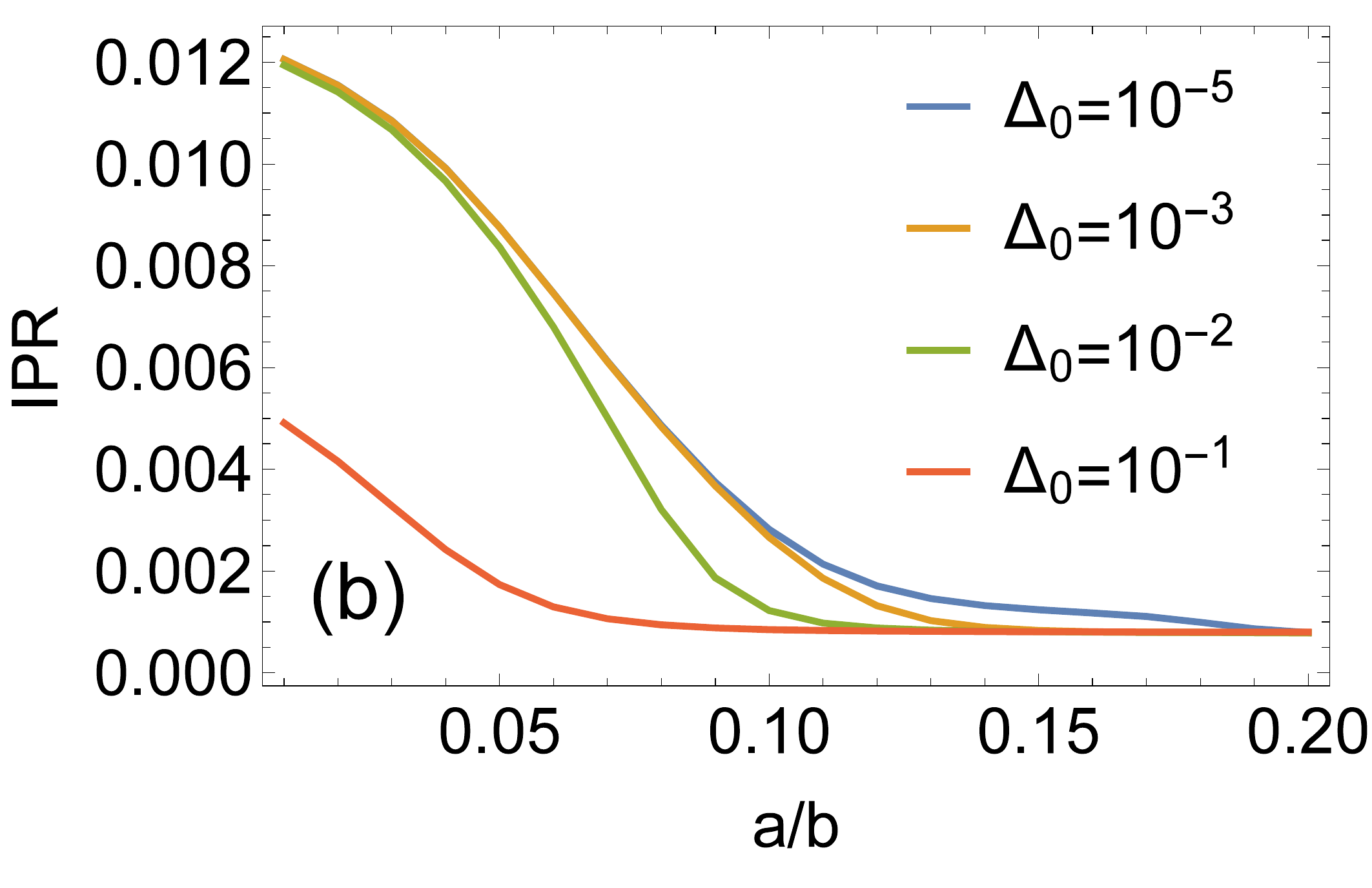} \\
\includegraphics[width=0.7\columnwidth, angle=0]{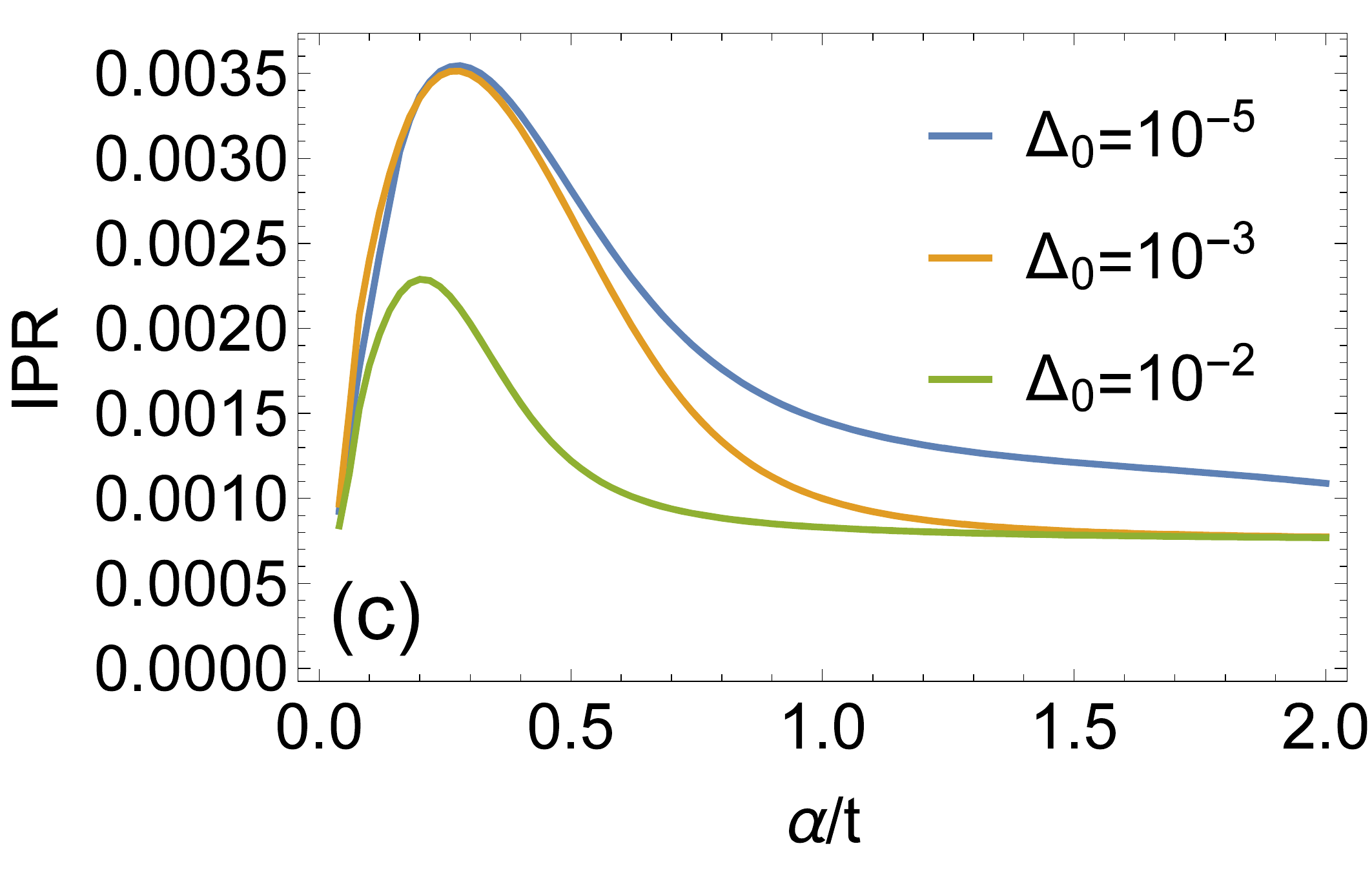}
\includegraphics[width=0.7\columnwidth, angle=0]{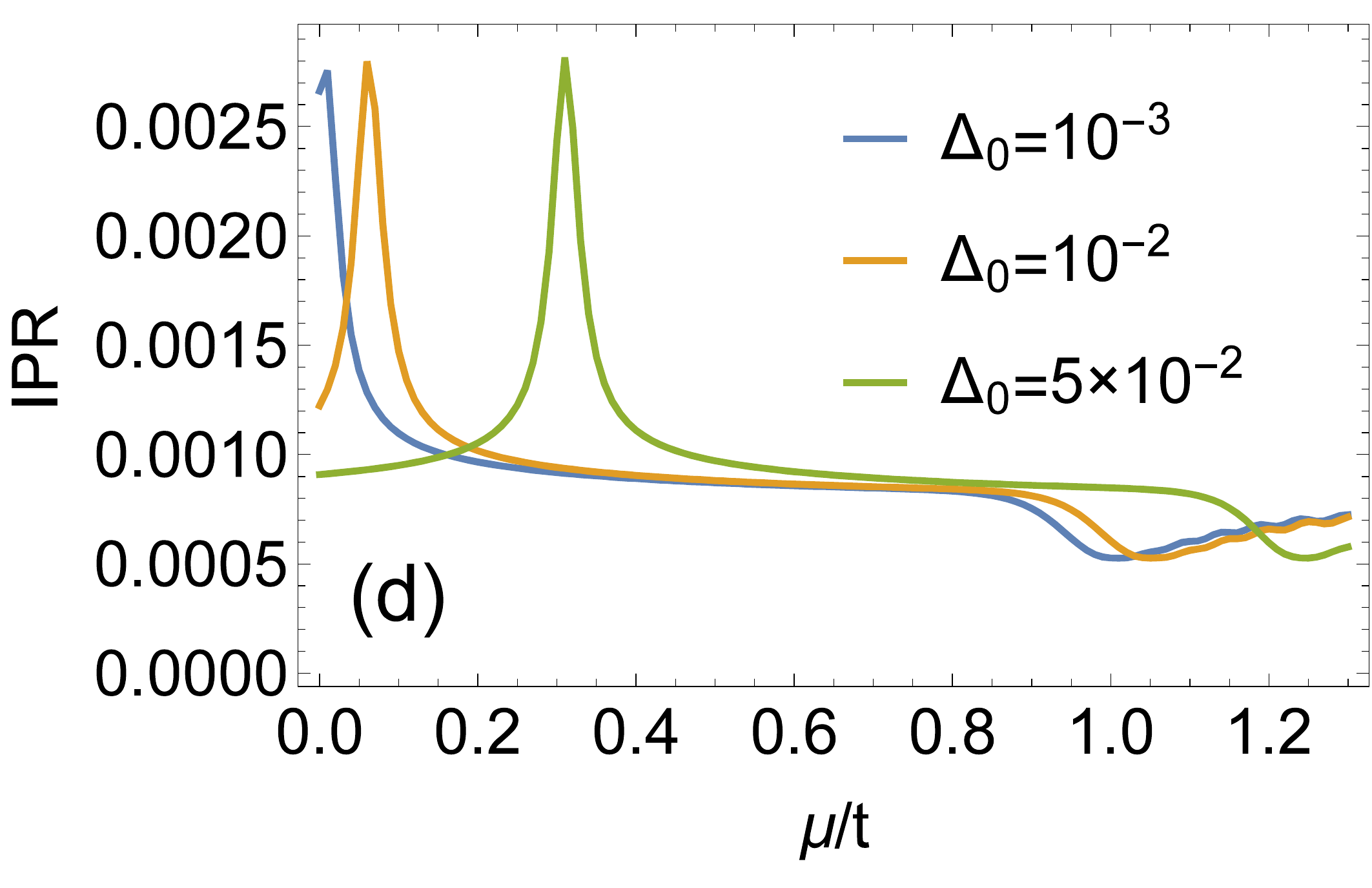}
\caption{The energy spectrum of a semi-elliptical wire as a function of $a/b$ (a) and the inverse participation ratio (IPR) of the subgap mode as a function of $a/b$ (b), $\alpha/t$ (c), $\mu$ (d). 
When the wire is squeezed enough, a two-fold degenerate level emerges within the energy gap. It corresponds to two modes which are spatially localized at the center of the wire.  We used the following set of parameters: $t=1$, $\mu=0$, $\Delta_0=0$, $\Delta_1=0.2 t$, $N=500$, $\alpha/t = 0.5 $ in (b), and $a/b=0.1$ in (c). In (d) it is $\alpha/t = 0.5 $ and $a/b=0.1$. }
\label{fig:locspe}
\end{figure}

By analyzing the energy spectrum, we observe that, when the wire is in the topological phase, and it is strongly squeezed, a Kramers pair of quasi-particle bound states nucleates inside the bulk energy gap, as shown in Fig.~\ref{fig:locspe} (a).

The performed analysis shows that a strong squeezing induces a strong curvature variation (corresponding to a rapid variation of the local polar angle $f_j$) 
at the wire's central point. Such curvature inhomogeneity produces a local attenuation of the effective strength of the spin-orbit interaction just at the point where the curvature assumes the largest value. As a consequence, the emerging in-gap modes can be attributed to a strong inhomogeneity of the topological phase around the point of maximum curvature. 

By reducing the wire length, these ingap states hybridize more and more with the zero energy states of the topological phase, which are localized at the wire ends.

The localization of the ingap states can be hestimated by determining the inverse participation ratio (IPR), which is defined for the $n-th$ energy level through the site dependent coefficients $\mathbf{u}_{n,i},\mathbf{v}_{n,i}$ of the corresponding wave function as $IPR_n = \sum_{i=1}^N (\abs{\mathbf{u}_{n,i}}^4+\abs{\mathbf{v}_{n,i}}^4)$.

As shown in Fig.~\ref{fig:locspe} ((b)-(d)), the increasing localization of the in-gap states due to a strong squeezing (or due to a reduction of the RSOC amplitude) leads to an increasement of the corresponding IPR.

By switching on a finite on-site superconducting pairing $\Delta_0$, the IPR becomes suppressed, due to the fact that the in-gap modes are absorbed into the continuum band, as it can be observed both in the region of high squeezing (see Fig.~\ref{fig:locspe}(b)) and at small Rashba SOC values (see Fig.~\ref{fig:locspe}(c)).

By varying the chemical potential, as shown in Fig.~\ref{fig:locspe}(d) for the representive case $\alpha/t = 0.5$ and $a=b = 0.1$, the localization undergoes a strong enhancement at the value of the chemical potential corresponding to $\alpha^{*} \approx 0$, which is larger at larger onsite pairing $\Delta_0$.

These peaks in the IPR behavior actually correspond to the localization of the in-gap states at the center of the semiellipse. Moreover, above the critical value of the chemical potential where the system undergoes a topological transitions into the trivial phase, the IPR has a sudden drop, which indicates a clear delocalization of the considered energy levels, due to their absorption in the bulk continuum.

\section{IV. Spin-triplet correlations}

For a curved profile, the superconducting ground state can exhibit spin-triplet correlations with nontrivial textures in space~\cite{nostroSC}, due to the curvature induced torque on the single electron spin momentum \cite{nostroPRB16}.

\begin{figure}[!ht]
 \includegraphics[width=0.49\columnwidth]{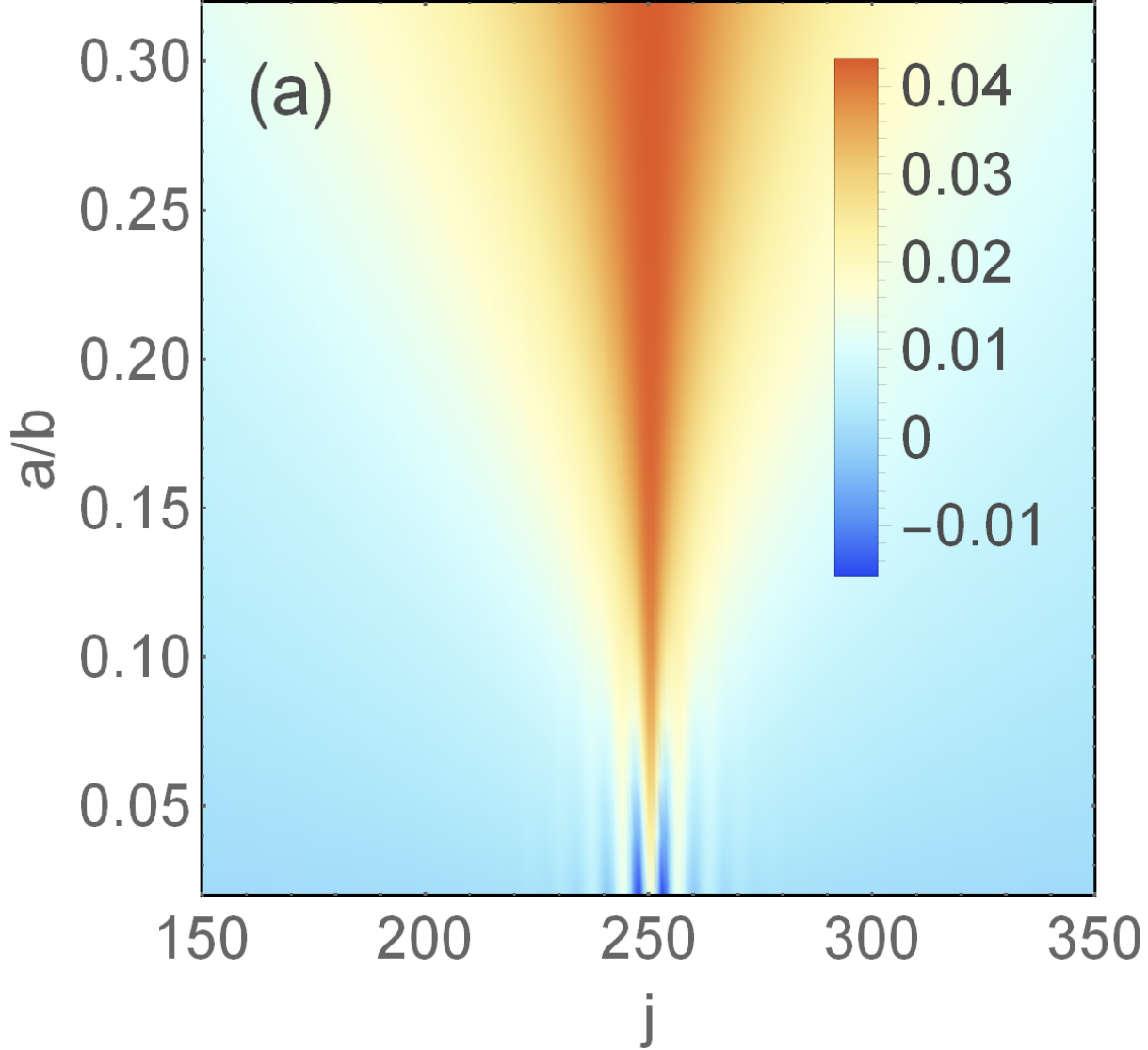} \includegraphics[width=0.49\columnwidth]{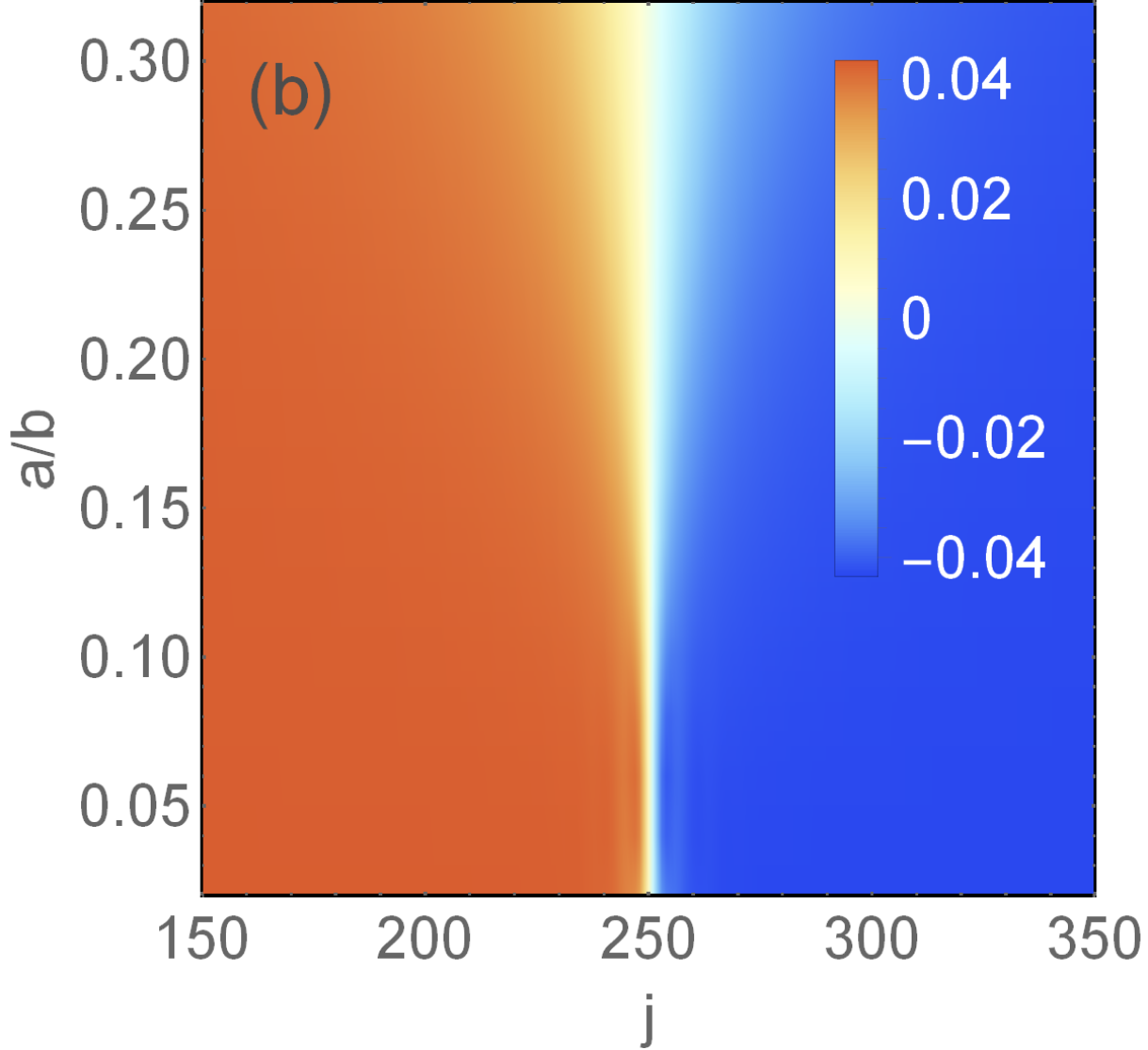} \\
 \includegraphics[width=0.49\columnwidth]{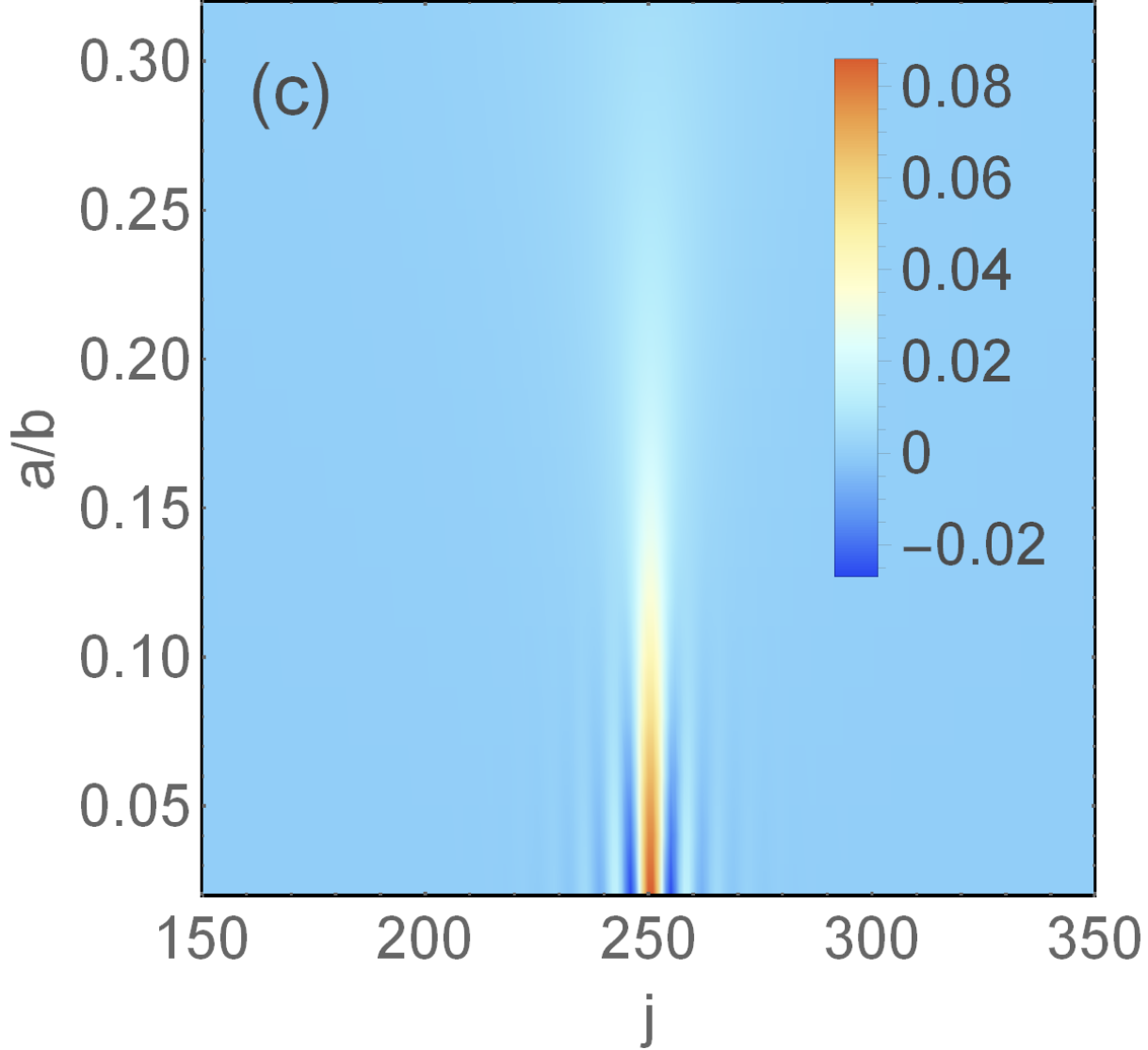} \includegraphics[width=0.49\columnwidth]{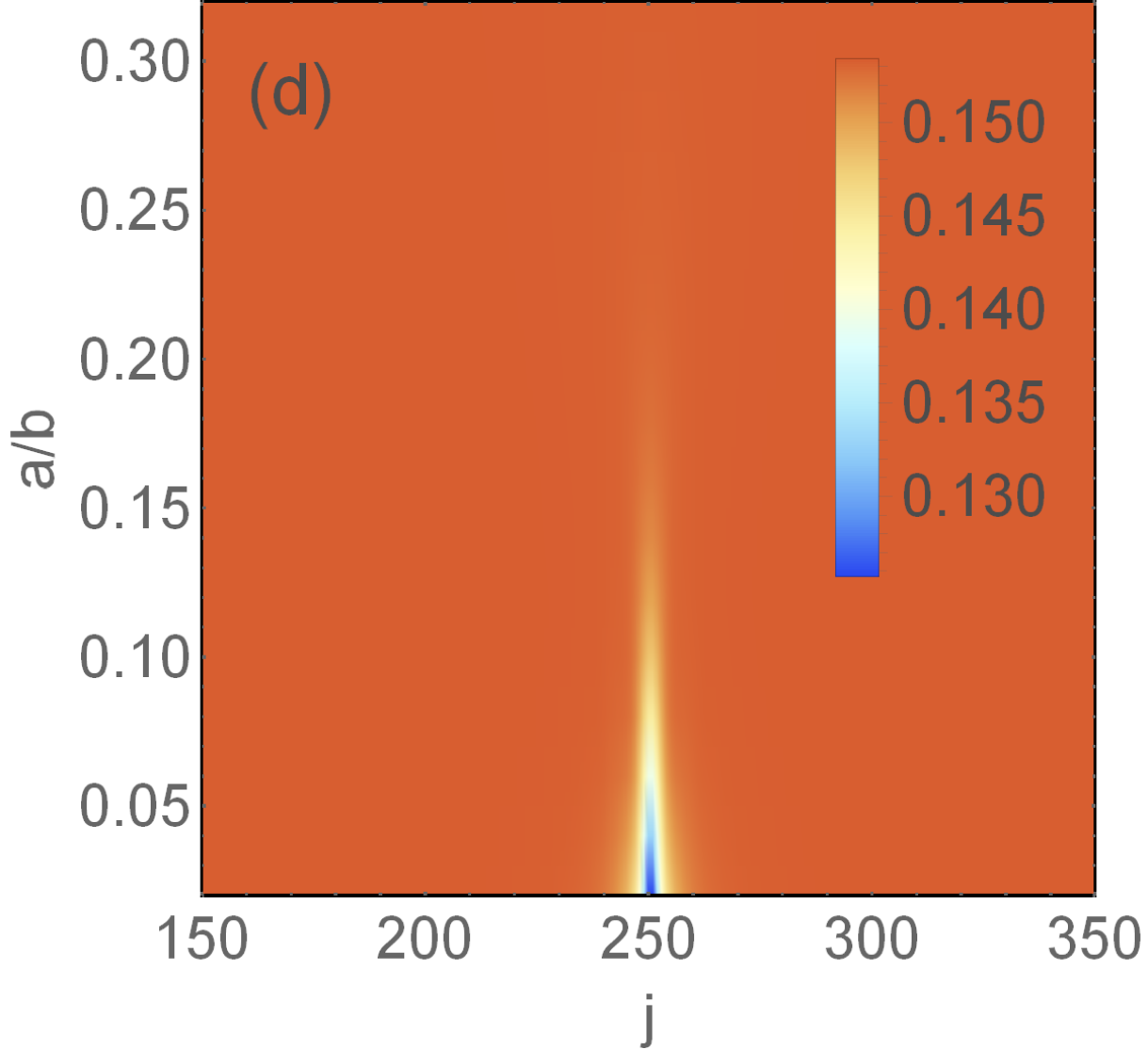}
\caption{ Spin-triplet correlations  $d_x$ (a), $d_y$ (b),$d_z$ (c) and local spin-singlet correlations $d_0$(d) of the superconducting wire as a function of the site position $j$ along the wire and of the semiaxes ratio $a/b$. We used $t=1$, $\mu=0$, $\Delta_0=0$, $\Delta_1=0.2 t$, $\alpha=0.5 t$ and $N=500$.
}
\label{fig:corr}
\end{figure}

Due to the inhomogeneity of the system, it is convenient to look at the spin-triplet correlators in real space $d^x_j = \frac{1}{2}(-\langle c_{j\uparrow} c_{j+1\uparrow}\rangle +  \langle c_{j\downarrow} c_{j+1\downarrow} \rangle)$, $d^y_j = \frac{1}{2 i}(\langle c_{j\uparrow} c_{j+1\uparrow}\rangle +  \langle c_{j\downarrow} c_{j+1\downarrow} \rangle)$ and $d^z_j = \langle c_{j\uparrow} c_{j+1\downarrow}\rangle +  \langle c_{j\downarrow} c_{j+1\uparrow} \rangle$. The behavior of the nearest neighbour spin correlations, derived by changing the local curvature, is reported in Fig.~\ref{fig:corr}.

As in the case when the spin-singlet pairing interactions are local ~\cite{nostroSC}, also in the non-local pairing case here considered, the superconducting correlations pattern gets modulated in amplitude and orientation due to curvature inhomogeneities.
In particular, when the semielliptical wire is strongly squeezed, the system behaves as composed by two regions with large curvature radius (thus nearly flat) separated by a central region with large curvature. We thus expect that superconducting correlations will be almost homogenous away from the wire center, with significant variations just around it.

Since spin-triplet correlations follow the RSOC vector field ~\cite{sigrist}, which is point by point perpendicular to the electron motion, in the case of a flat wire the $\mathbf{d}$ vector has homogeneous amplitude and direction, and if electrons are assumed to move along the $x-$axis, $\mathbf{d}$ vector is oriented along the $y-$axis direction. After bending the nanowires, a local rotation of the  $\mathbf{d}$ vector occurs, such that additional components of the  $\mathbf{d}$ vector emerge.

In particular, bending the superconducting wire leads to a finite $d^x_j$ component. It has a smooth space evolution and reaches its maximum values around the point of the largest curvature. When the squeezing becomes strong, and curvature starts to be increasingly inhomogenous, the maximum value of $d^x_j$ correlations strongly increases at the center of the wire, resulting then suppressed a lot elsewhere. At very hard squeezing, oscillations of the sign of these correlations appear around the point of maximum curvature (Fig.~\ref{fig:corr}(a)).

Conversely, the $d^y_j$ component, which is the only finite one in the case of a flat wire, smoothly changes sign at the wire's center when we start bending the wire. In the strong squeezing limit, it undergoes an abrupt, step-like spatial variation at the point where the curvature strongly changes (Fig.~\ref{fig:corr}(b)).

Since geometric curvature acts as a torque field on the single electron spin, pushing it out of the plane where the bending occurs \cite{SPIN, nostroPRB16}, the increasing squeezing allows $d^z_j$ correlations to rise around the largest curvature point. 
Consequently, stronger the squeezing, more significant is the amplification of these correlations at the wire's center, which is accompanied by rapid oscillations of their sign (Fig.~\ref{fig:corr}(c)). 

We can thus conclude that, for strong squeezing, at the point of maximum curvature, the $d^z_j$ component tends to dominate over all the others, in such a way the $\mathbf{d}$ vector tends to be locally pinned along the $z$ direction. Around this point, a strong squeezing causes the $\mathbf{d}$ vector to wind around the $y$ axis.

For completeness, in Fig.~\ref{fig:corr}(d), the non local singlet correlation function $d^0_j = \langle c_{j\uparrow} c_{j+1\downarrow}\rangle - \langle c_{j\downarrow} c_{j+1\uparrow} \rangle$ is also shown. Non local singlet correlations are significatively affected by the inhomegeneity of the curvature only at hard squeezing $a/b\ll 1$, when they become suppressed by the strong curvature.

\section{V. Josephson effect} 
Once clarified the effects of curvature on the energy spectrum and on superconducting pair correlations of semiconducting nanowires in the DIII topological phase, the next step is to understand the role of curvature in the Josephson effect between  DIII TSCs.
The case of flat geometry has been already explored, considering Josephson junctions realized by interfacing a DIII TSC to a $s$-wave SC, or two TSCs coupled via a conventional SC \cite{chung2013,Liu2014,Gong2016}. Compared to the flat case, bent nanowires of DIII TSCs offer novel ingredients to the Josephson effect, provided in particular by the curvature induced extra in-gap states as well as by the presence of additional curvature induced spin-triplet superconducting correlations. 

Nevertheless, geometric curvature has been already recognized as a potential tool for the possible manipulation of the Josephson effect, both in the case of juctions based on non-topological SCs \cite{RSOCsplitters}, as well as in junctions of TSCs \cite{Spanslattprb2018}.
In the case of non-topological Josephson junctions, geometry plays an important role when the superconducting leads are connected via a spin-orbit coupled weak link. Indeed, SOC rotates the electron spin around an axis fixed by the electron momentum and the electric field responsible for the SOC. This axis can be geometrically manipulated if the weak link contains a SO coupled bent nanowire \cite{RSOCsplitters}. In such a case, the weak link acts as a ``spin splitter" of the spin states of Cooper pairs tunneling through the link, thus transforming the spin-singlet Cooper pairs into a coherent mixture of singlet and triplet states. The resulting interference pattern between channels with different spin character allows for electrical and mechanical control of the Josephson current between the two singlet SCs, such that the Josephson current results to be a periodic function of the bending angle of the nanowire \cite{RSOCsplitters}. 

A geometric tunability of the Josephson effect has been also demonstrated within the context of topological junctions involving one-dimensional time reversal symmetry broken SCs (chiral TSCs) \cite{nostroJoseph,Spanslattprb2018}. In this case, it has been found that a geometric offset angle between two flat wires can generate an anomalous contribution to the Josephson current, i.e. the flow of a supercurrent in the absence of  external phase bias. A proportionality relation between the local current density and the local curvature for a single curved wire has been also established \cite{Spanslattprb2018}. 
Interestingly, when juntions of chiral topological nanowires have strong SOC inhomogeneities, in particular when the RSOC of the two topological superconducting leads may change abruptly the sign, three topological phases protected by chiral symmetry are supported, and the application of a superconducting phase gradient over the interface of the TSCs reveals a $\pi$-junction behavior \cite{finlandese}. 

\begin{figure}[!ht]
\includegraphics[width=0.48\columnwidth]{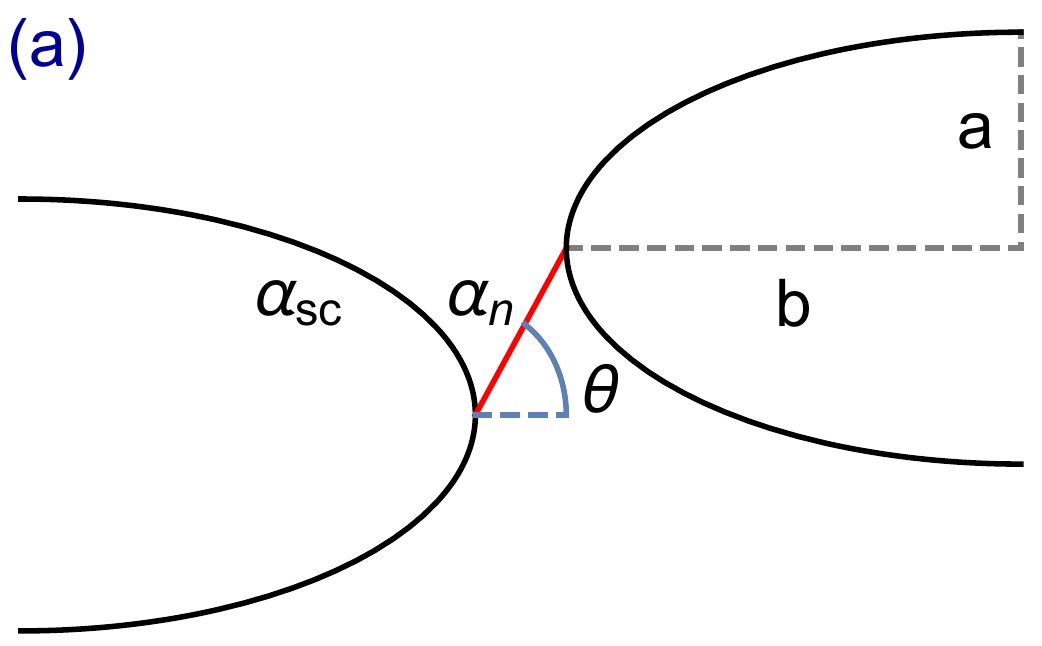} \includegraphics[width=0.48\columnwidth]{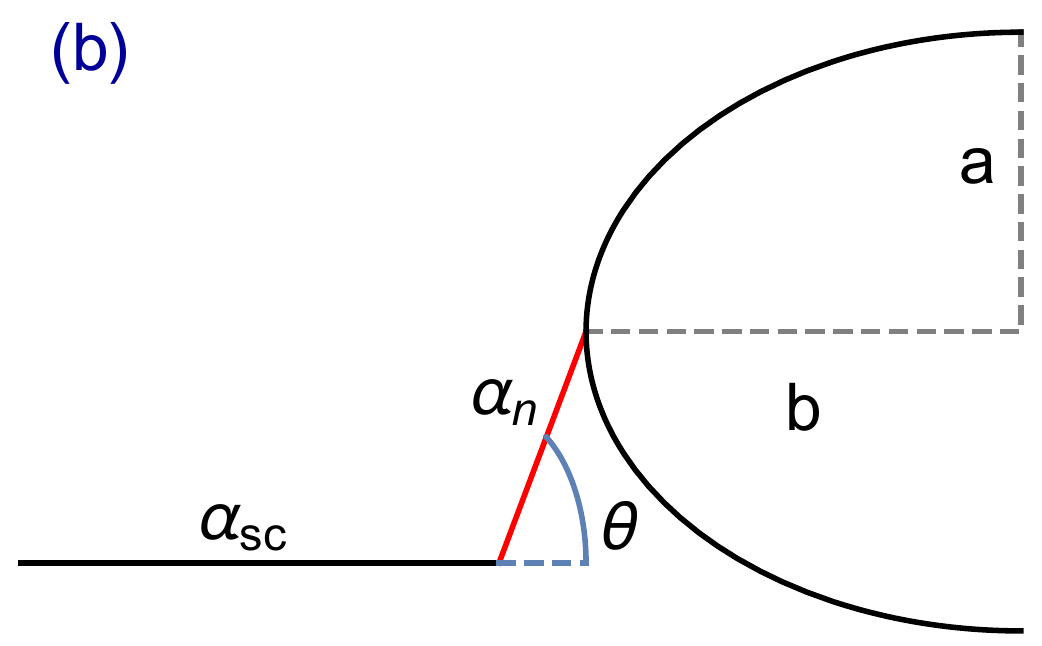}
\caption{Scheme of the considered junctions. In configuration (a), a semiconducting wire (red wire) connects the centers of two bent superconducting wires (black wires). On the other side, in configuration (b), the normal semiconducing wire connects the center of a superconducting bent wire with the edge of a superconducting straight wire. Both superconductors are assumed DIII TSCs. The normal wire has $N_{n}$ sites and the superconducting wire $N_{s}$. The superconducting wires are of the same length.}
\label{fig:scpic}
\end{figure}

Motivated by this rich scenario, we have considered two possible configurations where two DIII TSCs are interfaced through a normal semiconducting nanowire, as schematically represented in Fig.~\ref{fig:scpic}.

The first case we have addressed (Fig.~\ref{fig:scpic} (a)) involves a semiconducting Rashba spin-orbit coupled straight wire connected at each end to the center of a semielliptical superconducting wire. The two curved superconducting wires are characterized by parallel semiaxes of equal length. When the two external leads are in the topological non trivial phase, at hard squeezing values, the semiconducting internal wire is expexted to allow for the interaction among the in-gap states of the two semiellipses, which are localized close to the contact points between the central wire and the superconducting leads. 
 
In the second configuration (Fig.~\ref{fig:scpic}(b)), we have considered a straight semiconducting wire which is connected to two superconducting wires: one of the two is straight, while the other one is semielliptically shaped.  
The left (straight) wire of the junction is oriented parallel to the $b$ semiaxis of the right semielliptical lead. This configuration, differently from the first one, allows to investigate the interference between the Kramers doublet of Majorana modes generated at the right end of the straight wire in the topological regime with the in-gap states nucleating at the center of the right semielliptical lead.

In both configurations, the two superconducting wires have the same length and the Josephson effect is investigated as a function of the angle $\theta$ formed by the central short semiconducting wire with respect to the direction of the major semiaxis of the semielliptical wire.

When a phase difference $\varphi$ is applied to the two superconducting wires, an equilibrium current across the junction arises, which is given by $J = 2e/\hbar \partial_\varphi E_{GS}$, where $E_{GS} = - \Tr{\boldsymbol{\Lambda}}/2$ is the system ground-state energy. 
The total Hamiltonian is invariant under time reversal symmetry and also with respect to the sign change of the phase $\varphi$ of the two TSC order parameters, so that the current $J$ is an odd $2\pi$-periodic function of $\varphi$.

We consider large sizes $N\gg 1$ so that the band contribution is negligible.

When the system is in the flat configuration (all nanowires are flat and lay along the same direction), and the weak link does not have SOC, only a spin-singlet channel contributes to the Josephson current. However, when the external superconducting leads of the junction are bent, spin-triplet correlations emerge which are expected to activate new channels for the current propagation, which can be then manipulated through the direction of the Rashba SOC in the weak link. 

The analysis shows that the squeezing of the elliptical wires produces a very large amplification of the Josephson current with respect to the straight case, even in the absence of SOC in the weak link.
The amplification is much stronger when the semiconductor couples two equally squeezed semielliptical wires (Fig. 4 (a) and (c)), both having curvature induced in-gap states, than the case when the coupling involves a Majorana doublet and in-gap states (Fig. 4 (b) and (d)). 
The current amplification depends on the phase difference $\varphi$, as the energy of the subgap states generated by the curvature inhomogeneity sensibly changes with $\varphi$.
We also find a clear dependence of the amplification effect on the amplitude of the on-site singlet pairing: larger is this paring smaller is the amplification of the Josephson current.  Actually, a larger local pairing reduces the energy gap in the energy spectrum,so that the in-gap states become absorbed into the continuum. Their role in the Josephson effect becomes thus suppressed, and this can explain the detrimental effect of the local pairing on the amplification of the Josephson current. This supports a key role by the in-gap states in tuning the amplitude of the Josephson current.

\begin{figure}[!ht]
\includegraphics[width=0.49\columnwidth]{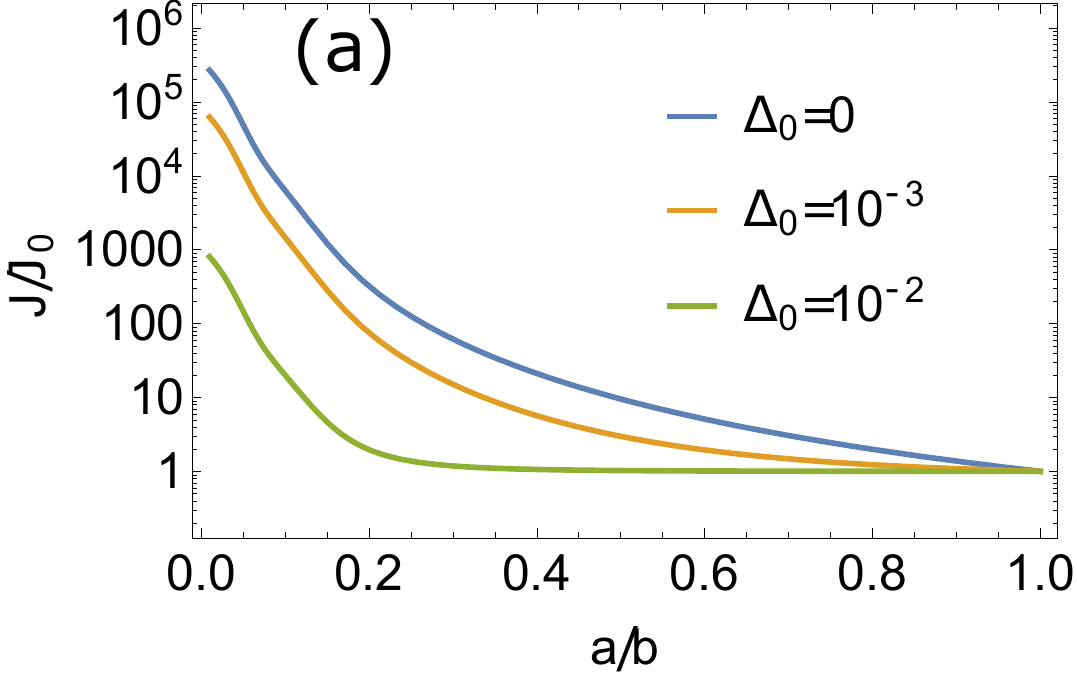} \includegraphics[width=0.49\columnwidth]{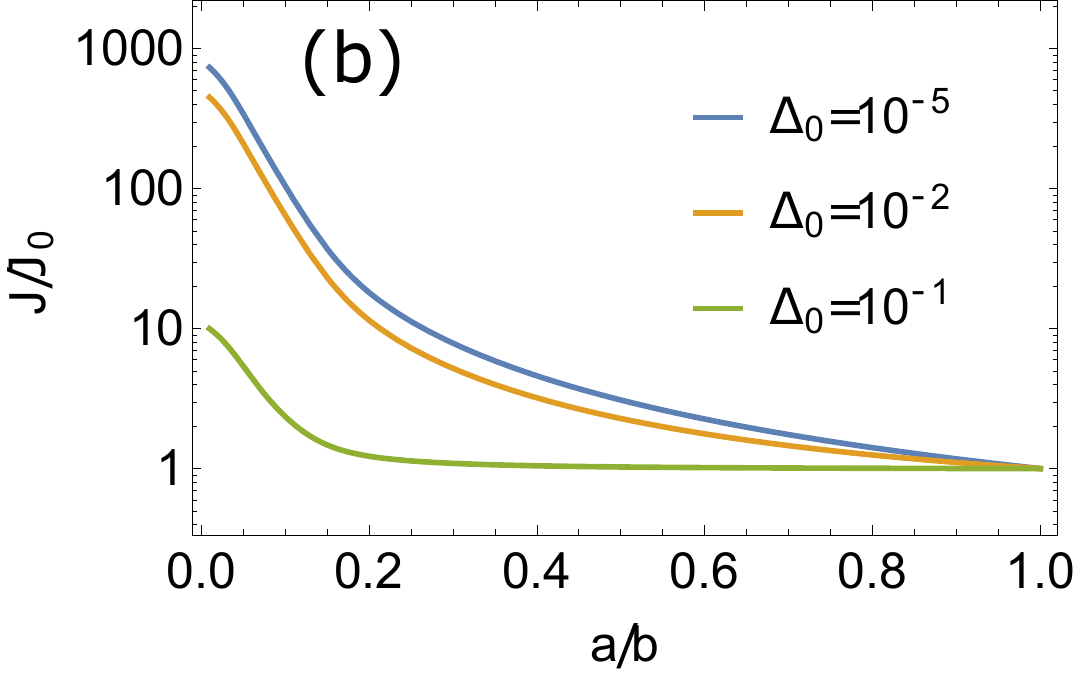}
\includegraphics[width=0.49\columnwidth]{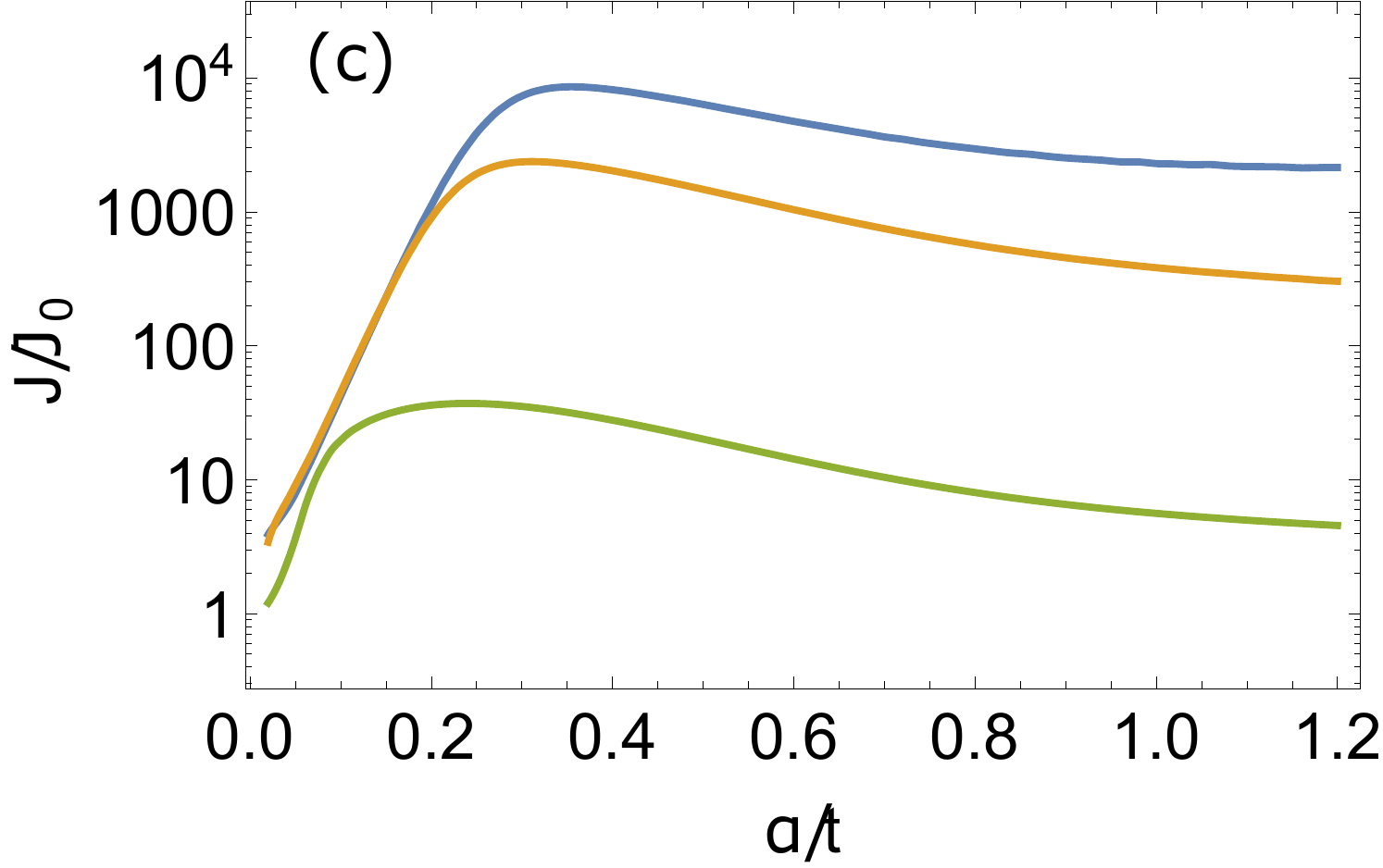} \includegraphics[width=0.49\columnwidth]{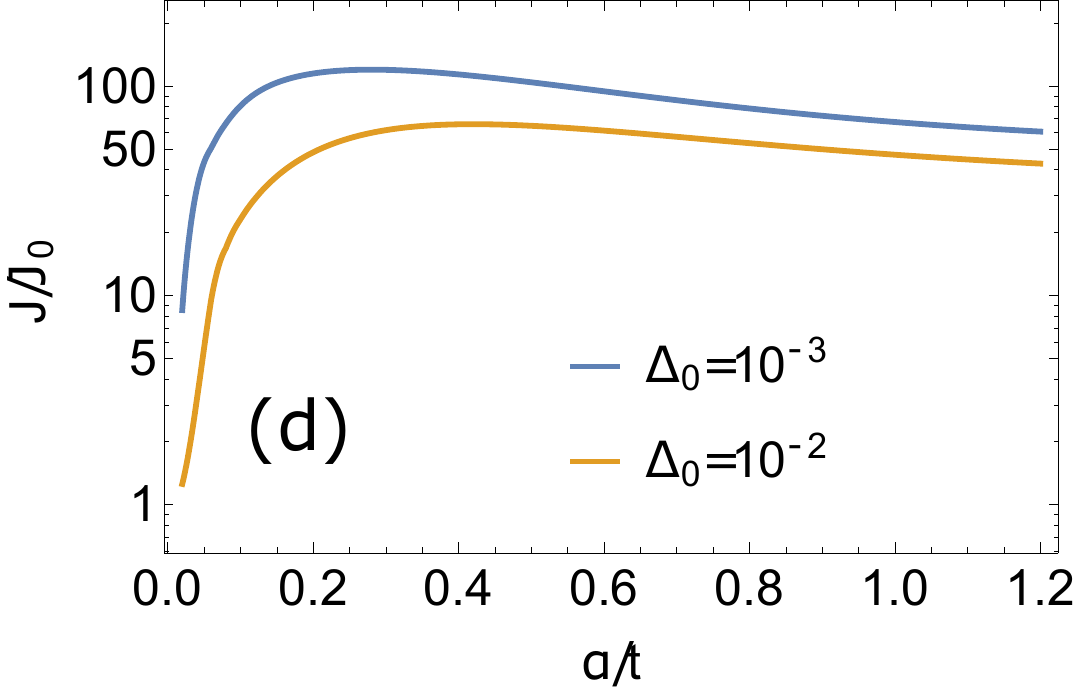}
\caption{Josephson current at $\varphi=\pi/2$ for the configuration in Fig. \ref{fig:scpic} (a) (left panels (a) and (c)) and for the configuration on Fig. \ref{fig:scpic} (b) (right panels, (b) and (d)) renormalized with respect to the maximum value $J_0$ of the homogeneous case ($a=b$), calculated as a function of the semiaxes ratio $a/b$ (in (a) and (b)) and of the RSOC amplitude $\alpha$ of the superconducting leads ((c) and (d)). We used $t=1$, $\mu=0$, $\Delta_1=0.2$, $\alpha_{sc}=0.5$, $\alpha_n=0$, $\theta=0$, $N_{sc}=500$, $N_n=3$, $a/b=0.1$.
}
\label{fig:jos_ampl_vsa_0}
\end{figure}

When the Cooper pairs contributing to the current have spin-triplet correlations, they undergo a spin rotation in the tunnelling across the junction due to the presence of the RSOC $\alpha_n$ in the semiconducting wire, allowing for a SOC controlled oscillatory critical current with a 0-$\pi$ transition in TSC/SOC-semiconductor/TSC Josephson junctions, both for class D TSCs \cite{liu15} and DIII TSCs \cite{haim19}.

In our case, we find a $0-\pi$ transition in both configurations (see Fig.~\ref{fig:jos_ampl_vsa_0}). It is modulated by the SOC amplitude $\alpha_n$ and orientation, the latter depending on the direction of the normal wire, specified by the angle $\theta$ (Fig. 4).

\begin{figure}[!ht]
\includegraphics[width=0.49\columnwidth]{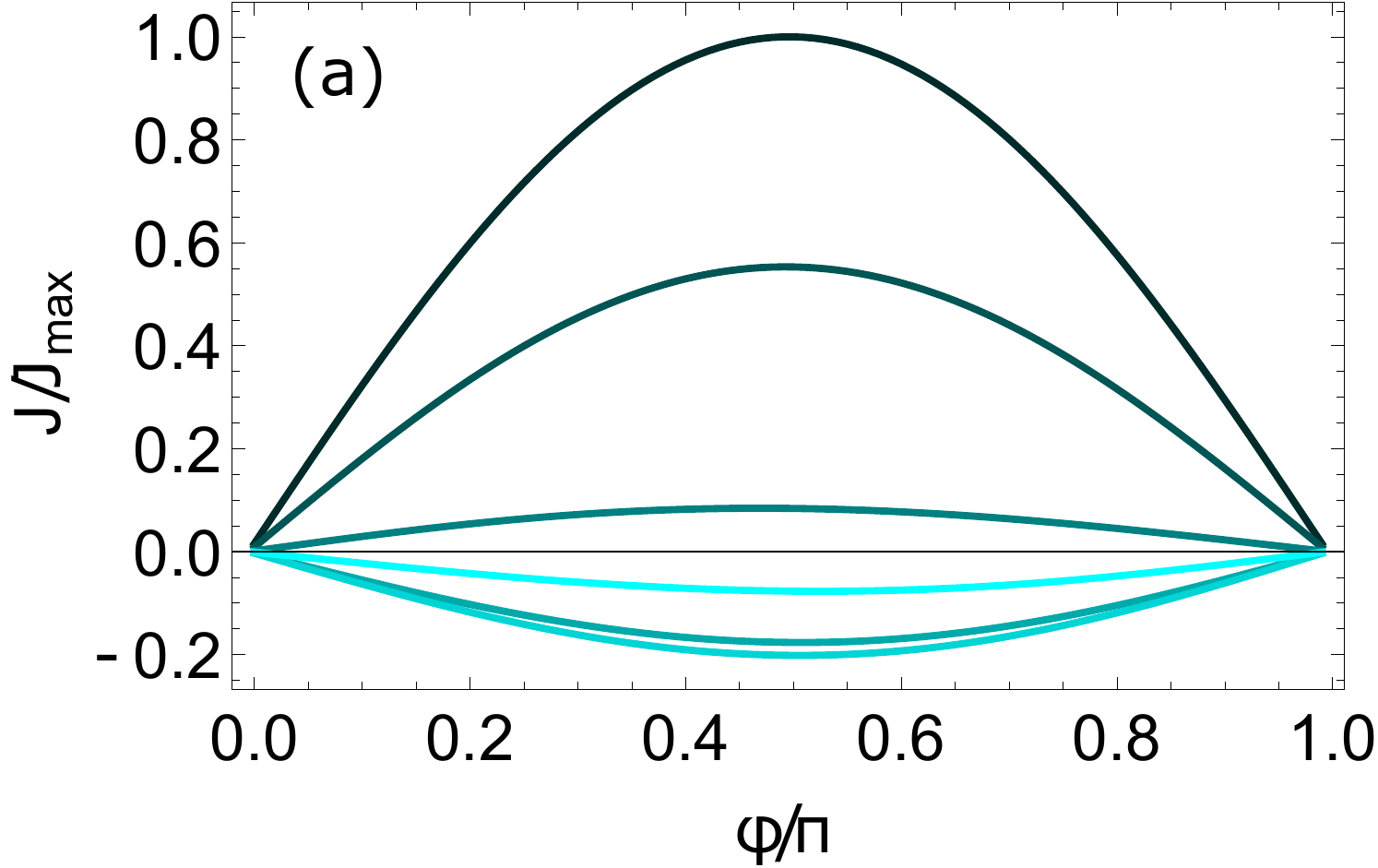} \includegraphics[width=0.49\columnwidth]{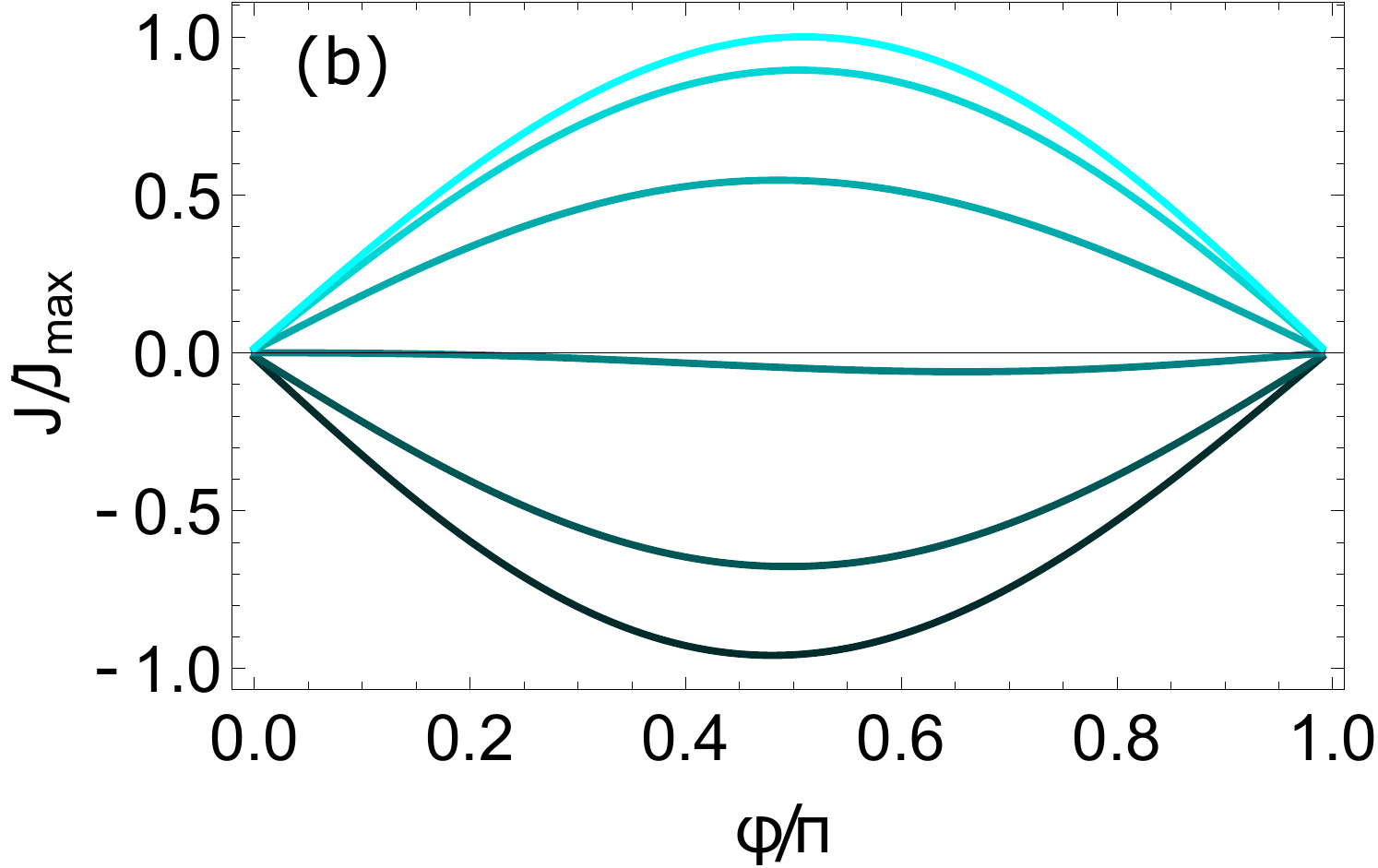}
\caption{ Josephson current as a function of the phase difference between the two TSCs for the configuration in Fig. \ref{fig:scpic}(a) (panel (a)) and for that in Fig. \ref{fig:scpic} (b) (right panel (b)) for the values $\alpha_n=0.,0.2,0.4,0.6,0.8,1.$ (darker to lighter) of the RSOC in the central non superconducting wire.  The currents are normalized with respect to their maximum values $J_{max}$. We used $\theta=\pi/4$ (a), $\theta=\pi/2$ (b), $t=1$, $\mu=0$,  $\Delta_0=0$, $\Delta_1=0.2$, $\alpha_{sc}=0.5$, $N_{sc}=500$, $N_n=3$, $a/b=0.1$.}
\label{fig:jos}
\end{figure}

In a flat configuration, for a time-reversal invariant TSC, there is a pair of Majorana zero modes with opposite spins at the boundary. The induced spin-triplet condensates are described by $(\ket{\uparrow\downarrow}+\ket{\downarrow \uparrow})/\sqrt{2}$ which is even under time-reversal operation (having assumed that the  Majorana  spins  are  parallel  or  anti-parallel  to  the $z-$  axis).  If $\theta=0$, the SOC  field direction is along $y-$axis, thus the electron spins are rotated in such a way that the spin-triplet states are changed to $(\ket{\nearrow\swarrow}+\ket{\swarrow\nearrow})/\sqrt{2}$, where $\ket{\nearrow}= \cos(\gamma/2) \ket{\uparrow}+ \sin(\gamma/2)e^{i\theta} \ket{\downarrow}$ is the rotated spin state, and $\gamma$ is the spin precession angle around the RSOC field~\cite{liu15}. Therefore the projection of the condensates is $(\bra{\uparrow\uparrow}+\bra{\downarrow \downarrow})(\ket{\nearrow\nearrow}+\ket{\swarrow\swarrow})/2=1$,  which is independent of the SOC. As a consequence, in this case the TSC/SOC semiconductor/TSC junction behaves like a conventional Josephson junction ~\cite{liu15}. 

By squeezing the semiellipse, additional spin-triplet components emerge, which can be manipulated by changing the direction of the normal wire, as well as by controlling the amplitude $\alpha_n$ of its RSOC.
In particular, for hard squeezing, the $d^z$ channel tends to be dominant in the TSC, and thus the projection condensate reads $(\bra{\uparrow\downarrow}+\bra{\downarrow \uparrow})(\ket{\nearrow\swarrow}+\ket{\swarrow\nearrow})/2=\cos \gamma$. 
Therefore, it does not explicity depend on the orientation angle $\theta$ of the semiconducting nanowire, but is fully determined by the amplitude of its SOC, such that the Josephson current can change sign by modulating the coupling $\alpha_n$. 

By analizing the Josephson current of the considered junctions at different orientation angles $\theta$, we actually observe that at strong squeezing a $\pi$-shift of the Josephson current occurs both in configuration (a) (Fig.~\ref{fig:jos}(a)) and in configuration (b) (Fig.~\ref{fig:jos} (b)) due to the amplitude modulation of the RSOC $\alpha_n$. 
 
\begin{figure}[!ht]
\includegraphics[width=0.49\columnwidth]{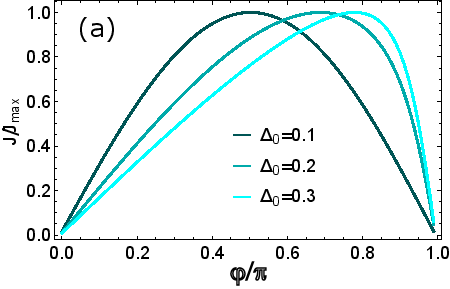}\includegraphics[width=0.49\columnwidth]{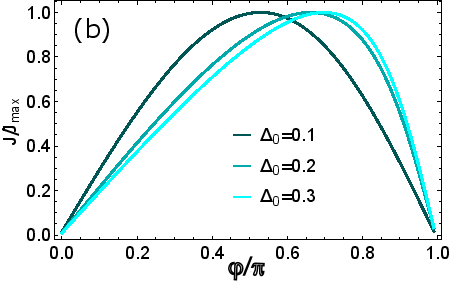}
\caption{Josephson current as a function of the phase difference between the two TSCs for the configuration of Fig. \ref{fig:scpic} (a) (panel (a)) and for the configuration of Fig. \ref{fig:scpic} (b) (panel (b)) for different values of a local singlet pairing $\Delta_0$, assumed to be the same in both TSCs. The currents are normalized with respect to their maximum value $J_{max}$. We used $\theta=0$, $\alpha_n=0$, $a/b=0.1$, $t=1$, $\mu=0$, $\Delta_1=0.2$, $\alpha_{sc}=0.5$, $N_{sc}=500$, $N_n=3$.
}
\label{fig:jos_2sh}
\end{figure}

\begin{figure}[!ht]
\includegraphics[width=0.49\columnwidth]{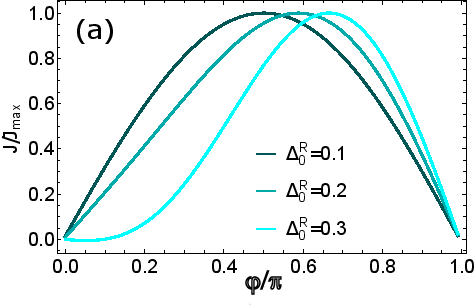}\includegraphics[width=0.49\columnwidth]{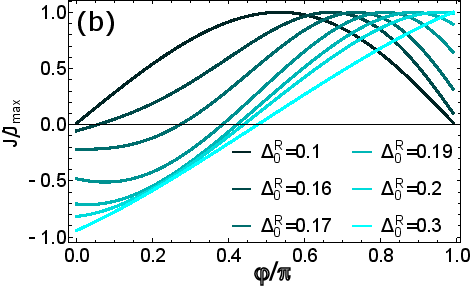}
\caption{Josephson current as a function of the phase difference between the two TSCs for the configuration of Fig. \ref{fig:scpic} (a) (panel (a)) and for the configuration of Fig. \ref{fig:scpic} (b) (panel (b)) in the case when the two TSCs have asymmetric superconducting local pairing values. We consider different values of the local pairing $\Delta^{R}_0$ in the right TSC, while fixing the value in the left TSC to $\Delta_0^{L}=0$. The currents are normalized with respect to their maximum value $J_{max}$. We used $\theta=0$, $\alpha_n=0$, $a/b=0.1$, $t=1$, $\mu=0$, $\Delta_1=0.2$, $\alpha_{sc}=0.5$, $N_{sc}=500$, $N_n=3$.
}
\label{fig:jos_2sh_as}
\end{figure}

Interestingly, the addition of a local pairing $\Delta_0\neq 0$, equal in both TSCs, gives rise to a second harmonic contribution, which is more significant as the local pairing becomes larger. Such a trend holds in both geometric configurations (see Fig.~\ref{fig:jos_2sh}).  

However, differently from configuration of Fig.\ref{fig:scpic}(a), in configuration of Fig. \ref{fig:scpic} (b) we find that, in the presence of a very small time reversal symmetry breaking source, it is possible to shift the current maximum at large $\varphi$ values, and simultaneously to generate an anomalous Josephson current contribution at zero phase by tuning the local pairing in the right TSC ($\Delta_R$) from the zero value up to the critical value $|\Delta_R|<2 |\Delta_1|$, above which the superconducting phase goes into the trival regime (see Fig.~\ref{fig:jos_2sh_as}).
Around this transition point, the Josephson current at $\varphi=0$ reaches its maximum absolute value, and the current-phase relation becomes almost linear.

\section{VI. Conclusions}
In summary, we have studied a time reversal symmetry invariant nanowire in the presence of spin-singlet superconducting pairing and RSOC. Such a system realizes a DIII TSC characterized by the presence of Majorana doublets at its ends. The analysis of the energy spectrum of this system in the topological phase shows that by strongly bending the wire in a semielliptical shape, finite energy localized states nucleate in the energy spectrum gap. Geometric bending of the wire also significantly affects the superconding pairing symmetry, allowing for the emergence of curvature induced spin-triplet correlations. As a consequence, the Josephson effect strongly depends on the geometric configuration of the junction: strongly squeezed semielliptically shaped superconducting nanowires can indeed allow for a huge amplification of the Josephson current, as well as for $0-\pi$ transitions and a finite flow of current at zero applied bias.  
Following the recent predictions about the realization of DIII topological phases, experimental verifications of the effects we found can be obtained by using strongly spin-orbit coupled semiconducting nanowires proximized by $d$-wave or $s\pm$-wave superconductors, like for instance InAs nanowire which are proximity coupled to high T$_c$ superconductors or to nodeless iron-based SCs. 
Finite size effects in the analized system have been found to be very important, as we report in the Appendix, where we show the possibility to turn the system from the DIII topological phase to a topologically trivial one by a suitable tuning of system length, geometric curvature and spin-orbit coupling.  
  
\section{VII. Appendix}
In the case of a system with non-homogeneous curvature, like for instance an elliptically shaped ring, novel Clifford pseudopsectrum methods can be employed to characterize the topology of the quantum phase~\cite{loring15}. For finite sizes $N$, the homotopy can be characterized by the real-space invariant which can be expressed in terms of the sign of the Pfaffian $W = \text{sign}(\text{Pf}(i (\mathbf{X} \boldsymbol{\Gamma} + \mathbf{H}'_{BdG})))$ where $\mathbf{X}$ is the position operator, and $\mathbf{H}'_{BdG}$ is the BdG matrix in the basis in which it is self-dual and imaginary~\cite{loring15}. 
In particular, this representation can be realized through the unitary transformation $\mathbf{H}'_{BdG}= \mathbf{U} \mathbf{H}_{BdG}\mathbf{U}^\dagger$ with $\mathbf{U}=\mathbf{U}_2\mathbf{U}_1$, being $\mathbf{U}_2=(1_{4N}+i \tau_x\otimes1_{N}\otimes \sigma_y)/\sqrt{2}$, and $\mathbf{U}_1=\frac{1}{\sqrt{2}}\left(
                                                                                                                             \begin{array}{cc}
                                                                                                                               1 & 1 \\
                                                                                                                               -i & i \\
                                                                                                                             \end{array}
                                                                                                                           \right)\otimes 1_{2N}
$.
Moreover $\boldsymbol{\Gamma}'=\tau_y$, and $\mathbf{X} = \tau_0 \otimes \text{diag}(x_j) \otimes \sigma_0$, with $-1/2\leq x_j \leq 1/2$.

\begin{figure}[!ht]
 \includegraphics[width=0.49\columnwidth]{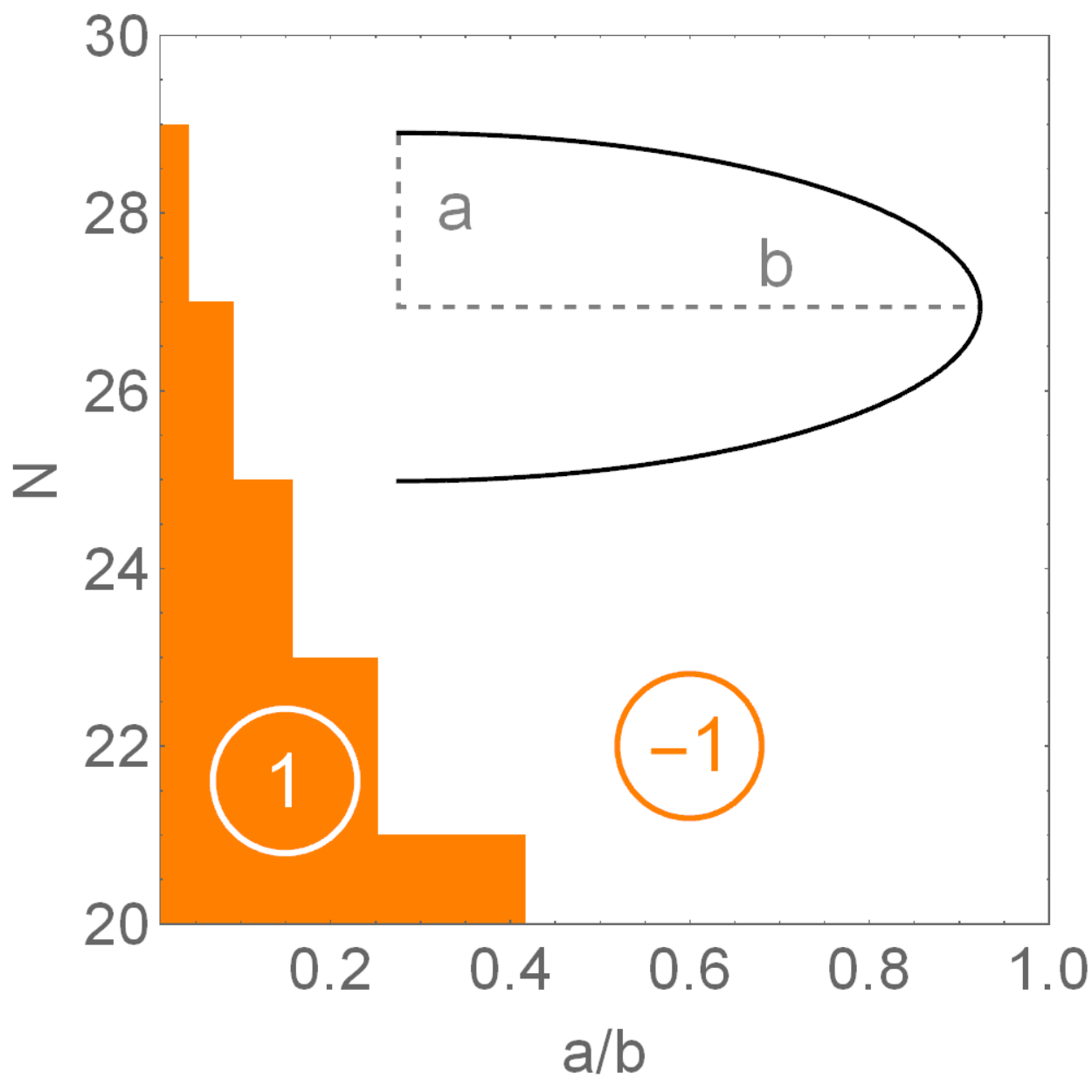} \includegraphics[width=0.49\columnwidth]{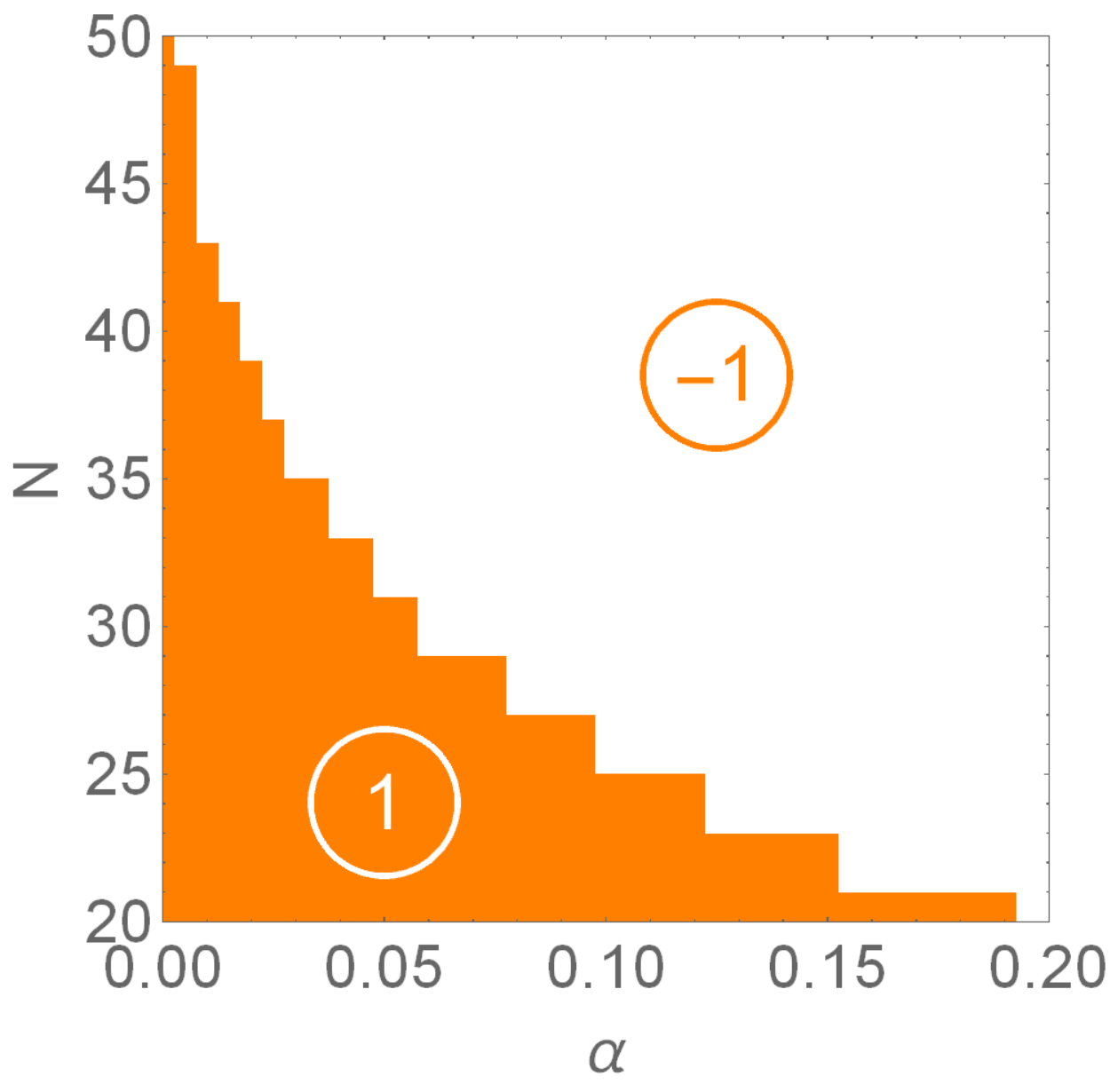}
\caption{Contour map of the invariant $W$ for a wire shaped as a half ellipse (inset in the left panel), as a function of the semiaxis length ratio $a/b$ (left panel), the RSOC $\alpha$ (right panel) and the size $N$.  We consider $\Delta_0 =\mu=0$, so that for a straight line the system is in its topological phase for $\alpha > 0$.  
We consider $t=1$ and $\Delta_1=0.2 t$. Moreover, in the left panel $\alpha=0.1 t$, and in the right panel $a/b=0.1$. The value of $N$ is changed in steps of $2$, while values of $a/b$ and $\alpha$ in steps of $0.005$.}
\label{fig:ti}
\end{figure}

We have thus analyzed the behavior of the invariant $W$ for the paradigmatic case of a superconducting wire having the shape of a half ellipse, characterized by the semiaxes length ratio $a/b$, and a RSOC of amplitude $\alpha$, as reported in Fig.~\ref{fig:ti}.

In the termodynmical limit, when $\Delta_0=\mu=0$ the system has a non trivial topological groundstate for $\alpha>0$, independently of the ratio $a/b$.
However, when the system has a finite size, the invariant $W$ is affected by the geometric parameter $a/b$, which can indeed drive a topological phase transition into a topologically trivial state in the limit of hard squeezing  $a/b \ll 1$. Shorter is the wire, larger is the effect of the curvature, leading to the suppression of the topological phase below a critical value of the semiaxes ratio $a/b$, which increases as the wire length decreases (Fig.\ref{fig:ti}, left panel). 

Analogous role is played by RSOC: in systems of finite size, there exists a critical value of RSOC below which the system is no longer in its topological phase. Such a critical value strongly increases at decreasing wire length (Fig.\ref{fig:ti}, right panel). 
We can thus conclude that 
 the topological quantum properties of the system are very sensitive to the system size, and are more significantly affected by the curvature as the system is shorter.

\end{document}